\makeatletter
\@ifundefined{@parse@version@dash}{%
\def\@parse@version#1{\@parse@version@0#1}
\def\@parse@version@#1/#2/#3#4#5\@nil{%
\@parse@version@dash#1-#2-#3#4\@nil}
\def\@parse@version@dash#1-#2-#3#4#5\@nil{%
  \if\relax#2\relax\else#1\fi#2#3#4 }
}{}
\makeatother

\documentclass[prl,twocolumn,superscriptaddress,showpacs,floatfix,longbibliography]{revtex4-2}
\usepackage{mathrsfs,braket}
\usepackage{amssymb, amsbsy, amsmath, latexsym, dsfont, array, layout, graphicx,mathrsfs,color,ulem,bm,soul}
\usepackage[colorlinks=true,citecolor=blue,urlcolor=blue,linkcolor=blue]{hyperref}

\begin{document}
\title{Observation of Higher-order Topological Bound States in the Continuum using Ultracold Atoms}

\author{Zhaoli Dong}
\author{Hang Li}
\email{lihang0125@gmail.com}
\author{Hongru Wang}
\author{Yichen Pan}
\affiliation{%
Zhejiang Key Laboratory of Micro-nano Quantum Chips and Quantum Control, School of Physics, and State Key Laboratory for Extreme Photonics and Instrumentation, Zhejiang University, Hangzhou 310027, China
}%

\author{Wei Yi}
\email{wyiz@ustc.edu.cn}
\affiliation{CAS Key Laboratory of Quantum Information, University of Science and Technology of China, Hefei 230026, China}
\affiliation{CAS Center For Excellence in Quantum Information and Quantum Physics, Hefei 230026, China}
\author{Bo Yan}
\email{yanbohang@zju.edu.cn}
\affiliation{%
Zhejiang Key Laboratory of Micro-nano Quantum Chips and Quantum Control, School of Physics, and State Key Laboratory for Extreme Photonics and Instrumentation, Zhejiang University, Hangzhou 310027, China
}%
\affiliation{%
College of Optical Science and Engineering, Zhejiang University, Hangzhou 310027, China
}%

\date{\today}

\begin{abstract}
Simulating higher-order topological materials in synthetic quantum matter is an active research frontier for its theoretical significance in fundamental physics and promising applications in quantum technologies.
Here we experimentally implement two-dimensional (2D) momentum lattices with highly programmable ability using ultracold $^{87}$Rb atoms.
Through precise control of experimental parameters, we simulate a 2D Su-Schrieffer-Heeger model with this technique, and observe the characteristic dynamics of corner and edge-bound states, where the corner state is identified as a higher-order topological bound state in the continuum.
We further study the adiabatic preparation of the corner state by engineering evolutions with time-dependent Hamiltonians.
We also demonstrate the higher-order topological phase transition by measuring both the bulk topological invariant and the topological corner state.
%
Our new platform opens the avenue for exploring the exotic dynamics and topology in higher synthetic dimensions, making use of the rich degrees of freedom of cold atoms systems.
%


\end{abstract}

\maketitle



{\it Introduction.}
Topological materials have attracted considerable interest due to their importance in fundamental physics and potential applications in quantum devices~\cite{PhysRevLett.45.494, RevModPhys.82.3045, RevModPhys.83.1057,Lu2014,RevModPhys.91.015006,RevModPhys.91.015005}. Recently, a new class of topological materials, dubbed higher-order topological insulators (HOTI)~\cite{science.aah6442,PhysRevB.96.245115,PhysRevLett.120.026801,PhysRevB.98.205147,PhysRevLett.119.246402,PhysRevLett.119.246401,sciadv.aat0346,PhysRevB.99.245151, PhysRevLett.123.256402}, has been proposed and verified on various physical platforms such as microwave resonators~\cite{Peterson2018}, photonics~\cite{,PhysRevLett.122.233902,PhysRevLett.122.233903,Noh2018, Mittal2019,Ota.19,Li2020,Zhang2020xu}, acoustics~\cite{Serra-Garcia2018,Ni2019,Xue2019,PhysRevLett.122.244301,Ni2020,Zhang2019,Zhang2019b,Xue2020,Zhang2020}, electric circuits~\cite{Imhof2018,PhysRevB.100.201406}, and solid materials~\cite{Kempkes2019}. Generally, a $d$-dimensional lattice can host ($d-n$)-dimensional states at its boundaries, protected by the crystalline symmetries, and referred to as $n_{th}$-order topological insulators,
For the most commonly studied HOTIs~\cite{science.aah6442,PhysRevB.96.245115,PhysRevLett.120.026801,PhysRevB.98.205147}, their characterization relies on the boundary states being embedded within the band gap, and distinctly isolated from other states. Whereas for some HOTIs~\cite{Hsu2016,PhysRevB.100.075120, PhysRevB.101.161116}, the corresponding boundary states are embedded within the bulk continuum yet
still localized. These states are known as the bound states in the continuum (BICs).

The BICs maintain the properties of bundary states while remaining degenerate with states in the bulk bands. Meanwhile, they can still be easily excited and detected without hybridizing with the bulk states, and can even acquire topological protection to become higher-order topological BICs~\cite{PhysRevB.101.161116,PhysRevLett.125.213901}.
In recent years, higher-order topological BICs have been extensively investigated across various systems, including the photonic waveguides~\cite{PhysRevLett.125.213901,Wang2021} and topological electric circuits~\cite{PhysRevLett.132.046601}.
Nevertheless, the demonstration of the higher-order BICs in ultracold atoms is still lacking,
wherein the multitude of highly controllable degrees of freedom offer rich opportunities. 

In this work, we report the experimental observation of HOTI and the associated higher-order topological BICs in ultracold  $^{87}$Rb atoms. This is achieved through our newly developed momentum-lattice platform, where 2D lattice Hamiltonians can be engineered in a programmable fashion in the momentum space of cold atoms.
Thanks to the flexible control over the lattice geometry and site-resolved hopping patterns and phases,
momentum lattices based on ultracold atoms have unveiled a wealth of quantum dynamic phenomena over the past decade, including exotic topological matter and transport~\cite{Meier2016a,2018_science,2019_npj,chiral_2024}, non-Hermitian quantum control~\cite{Gou2020, Liang2022}, interplay of mobility edges and interaction~\cite{PhysRevX.8.031045, PhysRevLett.126.040603, PhysRevLett.129.103401}, as well as correlated dynamics in frustrated geometries~~\cite{hang2022prl,Ma2023nc,zhang2023prl}.
While all these studies focus on one-dimensional or quasi-two-dimensional lattices, extending the momentum-lattice technique to higher dimensions is a highly desired yet challenging goal. Our implementation of the 2D square momentum lattice is hence a significant step toward achieving this goal.
%

Based on our latest technical progress, we implement the 2D Su-Schrieffer-Heeger (SSH) model and observe the characteristic dynamics of corner and edge bound states.
Importantly, the corner state is identified as the higher-order topological BIC of the model.
We experimentally confirm this by studying the adiabatic preparation of the corner state, and by
probing the higher-order topological transition of the system through measurement of the bulk topological invariant and the emergence of the BICs.

\begin{figure}[!t]
	\centering
	\includegraphics[width=1\linewidth]{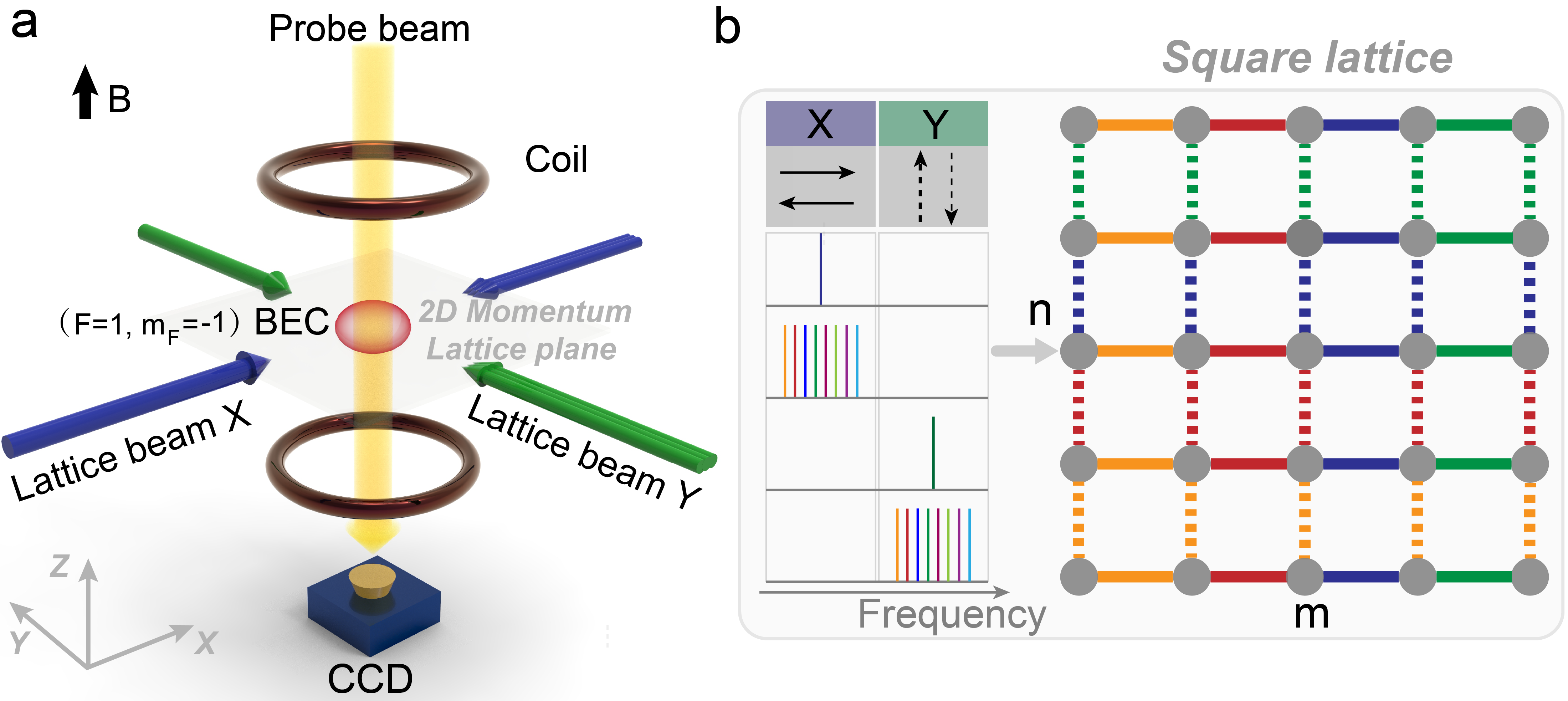}
\caption{\textbf{Schematic of constructing the 2D momentum lattice.} \textbf{a.} Schematics of constructing the 2D momentum lattice. Two sets of lattice beams are applied along the $x$ and $y$ directions to couple the momentum states in both directions.
\textbf{b.} (Left) The radio frequency spectrum applied for generating the momentum lattice in the $x$ and $y$ directions. (Right) Illustration of the generated 2D square lattice, bonds with the same color and line shape share the same hopping rate, as they undergo the same two-photon process.
        }
	\label{fig1}
\end{figure}

\begin{figure*}[t]
	\centering
	\includegraphics[width=0.95\linewidth]{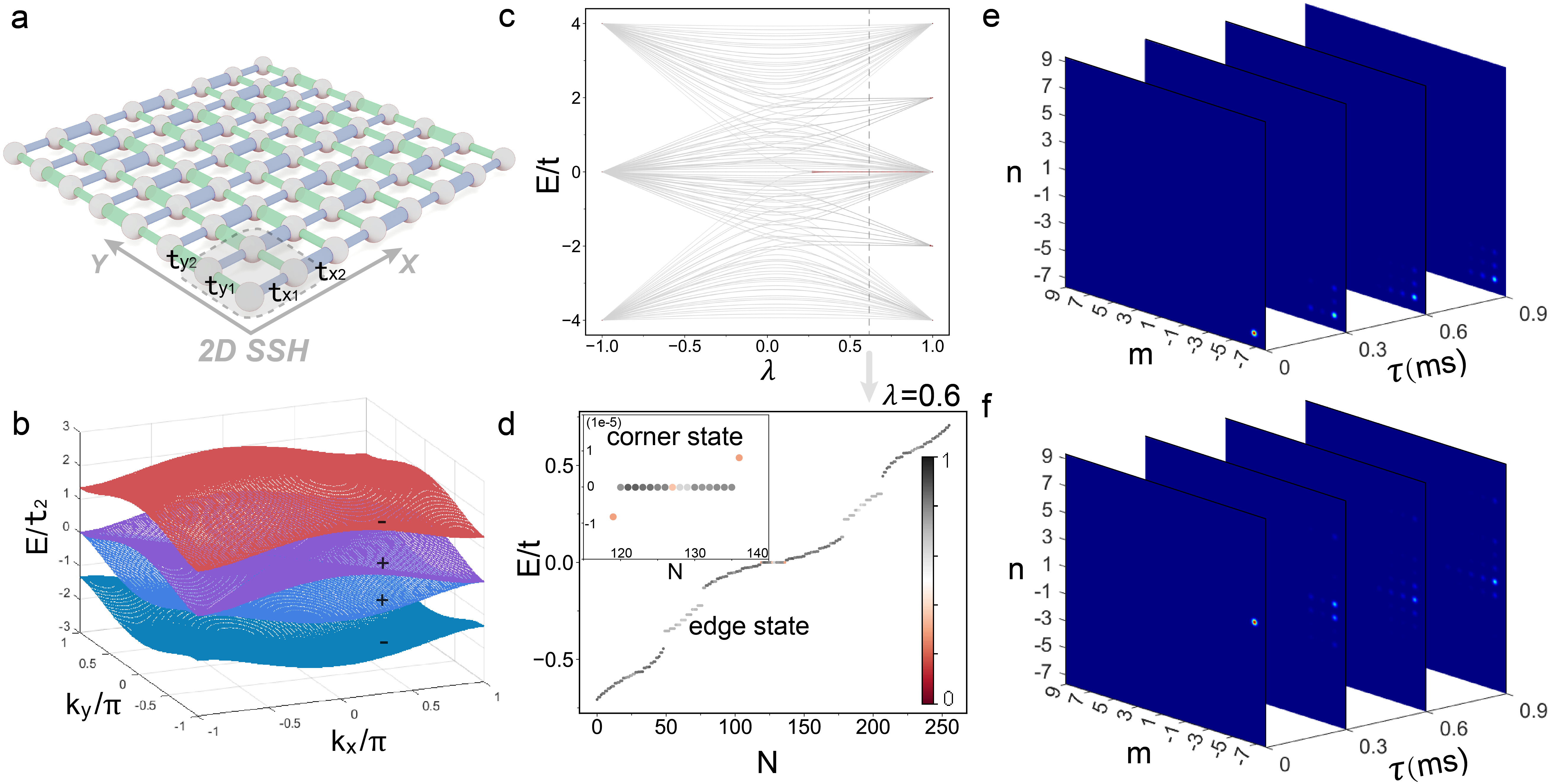}
	\caption{\textbf{The dynamics of corner and edge bound-states.} \textbf{a.} Illustration of the 2D SSH model, $t_{x1/y1}(t_{x2/y2})$ are the intracell (intercell) hopping rates between adjacent sites along the $x$ and $y$ directions. \textbf{b.} Bulk band spectrum in the higher-order topological phase with $\lambda=$0.5.  \textbf{c.} The quasienergy spectrum of the 2D SSH model in various topological phases. The zero-dimensional corner states are shown in red in the topologically nontrivial region. \textbf{d.} The quasienergy spectrum of the 2D SSH model with $\lambda$=0.6. There are edge states in the band gap. The zero-energy corner states are embedded within the zero-energy bulk states, that is, within a continuum. \textbf{e/f.} Experimental results of the corner/edge-state dynamics up to $\tau$=0$\sim$0.9ms.
		}
        \label{fig2}
\end{figure*}

{\it Implementing two-dimensional momentum lattice.}
We begin with a brief summary on the engineering of the 2D momentum lattice.
Building upon the state-of-the-art 1D momentum-lattice technique, we created a 2D lattice by applying two sets of perpendicular Bragg lasers on a $^{87}\textrm{Rb}$ Bose-Einstein condensate (BEC) with $4\times10^{4}$ atoms~\cite{2018_JOSAB,2021_npj,Dong2024,Liang2024}, as shown in Fig.~\ref{fig1}({\bf a}). The lattice lasers separately couple discrete momentum states of $\left|F=1, m_F=-1\right\rangle$ in ground-state hyperfine manifold along the $x$ and $y$ directions, giving rise to a 2D square lattice configuration. The discrete momentum states of this 2D square lattice are denoted as $p=(2m\hbar k_x+2n\hbar k_y)$, where the wave vectors $k_{x,y}=2\pi/\bar{\lambda}_{x,y}$ (with $\bar{\lambda}_{x,y}=794.7$nm), and $m,n\in \mathbb{Z}$ label the synthetic lattice sites [see Fig.~\ref{fig1}({\bf b})]. Hoppings between neighbouring momentum states along the $x$ or $y$ direction are realized by the corresponding two-photon Bragg processes, enabling a programmable control over the tunnelling terms of the synthetic 2D tight-binding model. Under our scheme, the nearest-neighbor couplings within the same column or row are induced by the same pair of two-photon Bragg process and therefore have the same coupling rates [see Fig.~\ref{fig1}({\bf b})]. To resolve the atomic population of the 2D momentum-lattice sites, we apply another probe laser from the top to perform the time-of-flight absorption imaging. More experimental details can be found in the Supplementary Material~\cite{supp}.


Based on the aforementioned experimental platform, we construct the 2D SSH model, as illustrated in Fig.~\ref{fig2}({\bf a}).
The model exhibits $C_{4v}$ symmetry, and each unit cell contains four sublattice sites. The implemented 2D momentum lattice is described by the tight-binding Hamiltonian
\begin{align}\label{Heff}
H_\text{eff}&=-\sum_{m,n\in \text{odd}}(t_{x1}c_{m+1,n}^{\dagger }c_{m,n}+t_{y1}c_{m,n+1}^{\dagger }c_{m,n})\nonumber\\
& \ \ \ -\sum_{m,n\in \text{even}}(t_{x2}c_{m+1,n}^{\dagger }c_{m,n}+t_{y2}c_{m,n+1}^{\dagger }c_{m,n})+H.c.,
\end{align}
where $c_{m,n}^{\dagger }(c_{m,n})$ is the creation (annihilation) operator for atoms on lattice site $(m,n)$, and $t_{x1/y1}=t(1-\lambda_{x/y})$ and $t_{x2/y2}=t(1+\lambda_{x/y})$ are the intracell and intercell hopping rate between adjacent sites along the corresponding lattice direction, respectively. In our experiments, we primarily focus on the scenario with $\lambda_{x/y}=\lambda$ and denote $t_{x1/y1}=t_{1}$ and $t_{x2/y2}=t_{2}$, where $\lambda\in (0,1)$.

The 2D SSH model, described by $H_{\text{eff}}$, permits two distinct topological phases that depend on the ratio of the intracell and intercell hopping rates~\cite{PhysRevLett.118.076803}. Figure~\ref{fig2}({\bf b}) displays the typical bulk energy spectrum of the topological nontrivial phase. Note that when we interchange the intracell and intercell hopping rates, both the topological trivial and nontrivial phases have the same bulk band structure~\cite{supp}. As shown in Fig.~\ref{fig2}({\bf c}), the topological phase transition occurs at $\lambda = 0$, where the bulk band gap of Fig.~\ref{fig2}({\bf b}) closes at the high-symmetry points. The quasienergy spectrum under the open boundary condition features two branches of edge states as the system transits from the topologically trivial phase to the nontrivial region. In the topologically nontrivial phase, the model possesses a second-order topological phase, characterized by a 2D Zak phase~\cite{PhysRevLett.118.076803} in the bulk, and zero-dimensional corner states at the boundary [marked in red in Fig.~\ref{fig2}({\bf c})].

Particularly, we show the energy spectrum of a 16$\times$16 square lattice for $\lambda$ = 0.6 in Fig.~\ref{fig2}({\bf d}). To characterize the localization of the eigenstates, we adopt the parameter $D_2=-\ln(\text{IPR})/\ln L$, where $\text{IPR}= \sum_{j}\left|\psi_{i}(j)  \right|^{4}$ is the inverse participation ratio of eigenstate $\psi_{i}(j) $ with eigenenergy $E_i$, and $L$ is the size of 2D lattice array. For a sufficiently large $L$, an eigenstate is considered extended when $D_2\sim 1$, and localized when $D_2\sim 0$. As labelled in the subfigure of Fig.~\ref{fig2}({\bf d}), there are several zero-energy corner states embedded in the continuum of the zero-energy bulk states. These zero-dimensional corner states are the high-order topological BICs, protected by the bulk topological invariant of the 2D SSH model, which is itself a HOTI.

\begin{figure*}[t]
	\centering
	\includegraphics[width=0.7\linewidth]{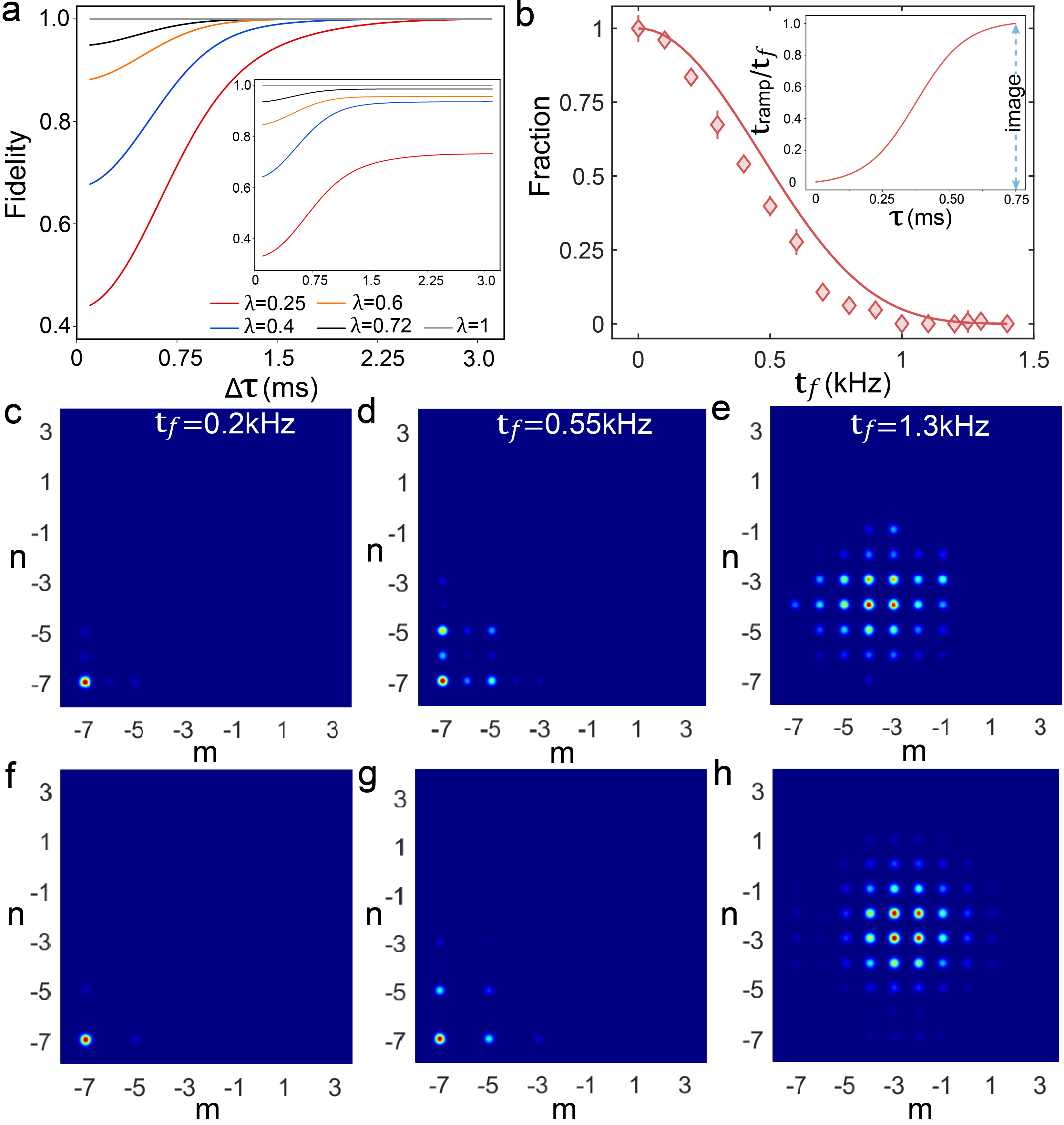}
	\caption{\textbf{Adiabatic preparation of the higher-order BICs.}  \textbf{a.} The simulated fidelity of adiabatically preparing the target BICs for various ramping periods $\Delta \tau$ and different $\lambda$ values in a 2D array of 16$\times$16. The inset shows The simulated fidelity of the 2D array of 15$\times$15. In both cases, $t_2$ is set to be $h\times$1.25kHz. \textbf{b.} Variation of the residual condensate fraction in the initial site for different $t_f$ values. The inset shows the variation of $t_1$ for the adiabatic ramping sequence \textbf{c-e.} The final states at the end of the adiabatic ramp, with $t_f=h\times$0.20(1)kHz, 0.55(2)kHz, and 1.30(2)kHz, respectively. \textbf{f-h.} The numerically simulated results, with $\tau=0.75$ms, for $t_f=h\times$0.2kHz, 0.55kHz, and 1.3kHz, respectively.
	}
        \label{fig4}
\end{figure*}

\begin{figure*}[t]
	\centering
	\includegraphics[width=0.8\linewidth]{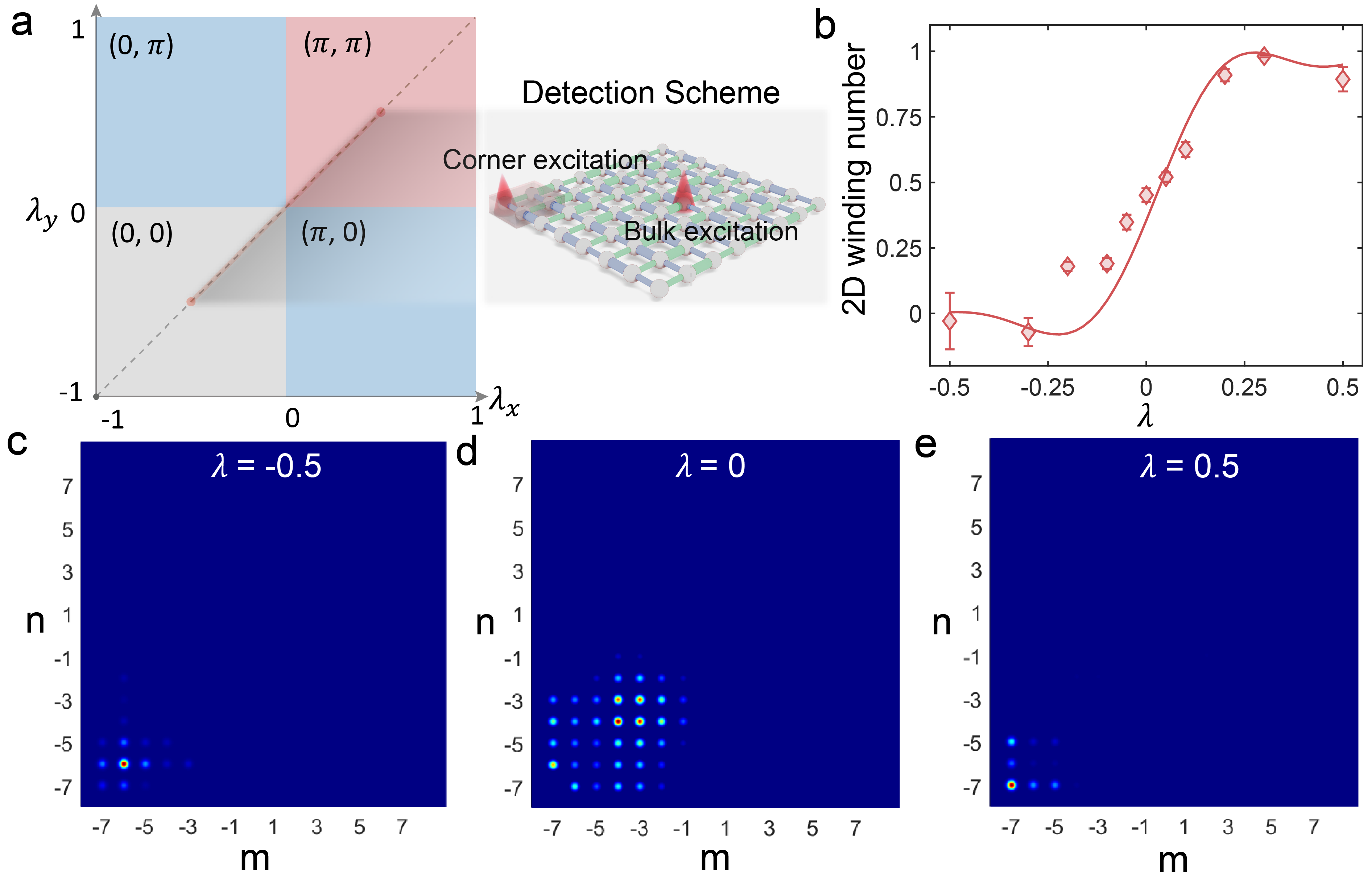}
	\caption{\textbf{Measuring the higher-order topological phase transition.}  \textbf{a.} The theoretical topological phase diagram with different $\lambda_x$ and $\lambda_y$, characterized by the 2D Zak phase. The subfigure illustrates our measurements of the phase transition through the corner-site and bulk-site excitations, respectively. \textbf{b.} The extracted 2D winding number varying with different $\lambda$. The dots are experimental data with $t=h\times 0.98(2)kHz$, and the solid lines are numerical simulations using $H_\text{eff}$. \textbf{c-e.} The experimental results following corner-site injection after $0.5$ms evolution with different $\lambda$ (hence different topological phases).
	}
        \label{fig5}
\end{figure*}

%

{\it Observing bound-states dynamics.} 
Notably, the corner states are protected by the $C_{4v}$ and chiral symmetries, so that they do not hybridize with the bulk states.
This gives rise to intrinsic corner bound-states dynamics that can be easily verified experimentally. To investigate this, we initialize the BEC at either a corner site or the midpoint of an edge, allowing us to examine the bound-states dynamics.
Figure~\ref{fig2}({\bf e}) shows the population at variable evolution times $\tau$, when atoms are initialized at the corner site. Here the large overlap with the corresponding zero-energy BIC restricts the wave function diffusion into the bulk, leading to the observed corner-localized dynamics.
Correspondingly, Fig.~\ref{fig2}({\bf f}) shows that the dynamics remains edge-localized when atoms are injected into the midpoint of the left edge, due to confinement of edge states along the 1D edge.



{\it Symmetry-protected adiabatic preparation of BICs.}
Leveraging the programmability of the 2D momentum lattice, the higher-order topological BICs can be
adiabatically prepared by implementing a symmetry-preserving time-dependent Hamiltonian.
We begin by populating the corner site, which is exactly the zero-energy BICs in the limit of hopping rates $t_1$=0 and $t_{2}=t(1+\lambda)$. Subsequently, we adiabatically ramp up $t_1$ from zero to $t_{f}$. The overall process is governed by the following time-dependent Hamiltonian
\begin{equation}\label{Hramp}
\begin{aligned}
H(\tau)&=\sum_{m,n\in \text{odd}}-t_{\text{ramp}}(\tau)(c_{m+1,n}^{\dagger }c_{m,n}+c_{m,n+1}^{\dagger }c_{m,n})\\& \ \ \  \sum_{m,n\in \text{even}}-t_2(c_{m+1,n}^{\dagger }c_{m,n}+c_{m,n+1}^{\dagger }c_{m,n})+H.c.
\end{aligned}
\end{equation}
where $\tau$ is the evolution time, $t_{\text{ramp}}(\tau)$ represents the time-dependent hopping rate.
Although the adiabatic preparation of the BICs might seem counterintuitive due to the absence of an energy gap,
the process is in fact protected by the $C_4$ symmetry, such that the bulk states nearby the BIC have vanishingly small overlap with the corner state and are not easily excited (see Supplemental Material).

Figiure~\ref{fig4}({\bf a}) presents the simulated fidelity of the adiabatic preparation, under various ramping periods $\Delta \tau$ and different $\lambda$ values in a 2D 16$\times$16 array. The inset of Fig.~\ref{fig4}({\bf a}) shows similar simulation results in a 2D 15$\times$15 array. They both show that the validity of this adiabatic preparation can be guaranteed for relatively long ramping time. The main difference between $L=16$ and $L=15$ mainly comes from the broken $C_4$ symmetry in the latter case. In the case with an odd $L$, the target BICs, localized in one corner and exhibiting broken symmetry, cause the finally fidelity to converge to a fixed value, which approaches unity as $\lambda$ approaches 1. While in the case of even $L$, the BICs are distributed across four corners under the $C_4$ symmetry, and show perfect fidelity for long enough ramping times~\cite{supp}.

%

In our experiment, the initial BEC wave packet can only be prepared in a single corner of the 2D array. Consequently, the actual adiabatic preparation process closely resembles the odd-$L$ case shown in Fig.~\ref{fig4}({\bf a}). We fix $t_{2}=h\times$1.25(2)kHz and adiabatically ramp up $t_1$ from zero to $t_{f}$ over a duration of 0.75ms, as shown in the inset of Fig.~\ref{fig4}({\bf b}). 
Figures \ref{fig4}({\bf c}) and \ref{fig4}({\bf d}) depict the experimentally prepared BICs at $\tau=0.75$ms with $t_{f}=h\times$ 0.20(1)kHz and 0.55(2)kHz, respectively. Figures \ref{fig4}({\bf f}) and ~\ref{fig4}({\bf g}) show the corresponding numerically simulated results for the two cases.
In either case, the final occupation is highly localized at even sites, consistent with the target
zero-energy corner state. In contrast, when $t_f$ becomes larger, as shown in Figs.~\ref{fig4}({\bf e}) and \ref{fig4}({\bf h}), the final occupation diffuses into the bulk, indicating the breakdown of the adiabatic preparation. The gradual breakdown of the adiabaticity is reflected in Fig.~\ref{fig4}({\bf b}), where we show the residual condensate fraction in the original corner site as a function of $t_f$.

{\it Measuring higher-order topological phase transition.}
To demonstrate the higher-order topological nature of the BICs, we match the measurements of the bulk topological invariants and the corner states, through bulk and boundary dynamics, respectively.

Theoretically, the higher-order topological phase transition is captured by the 2D Zak phases~\cite{PhysRevLett.118.076803}. In Fig.~\ref{fig5}({\bf a}), we plot the topological phase diagram characterized by the Zak phases. Experimentally, we probe the topological phase transition along the diagonal of the phase diagram Fig.~\ref{fig5}({\bf a}) by measuring the averaged 2D mean chiral displacement. More explicitly, we define the time-averaged mean chiral displacement as 
\begin{equation}\label{2Dwinding}
\upsilon _{2d}=\frac{1}{\bar{\tau}}\int_{0}^{\bar{\tau}}\bar{P}_{\mathcal{C}}(\tau)d\tau,
\end{equation}
where $\bar{P}_{\mathcal{C}}(\tau)=\left<\psi _{0} \right|e^{iH_\text{eff}\tau}\mathcal{C}e^{-iH_\text{eff}\tau}\left|\psi _{0} \right>$, $\mathcal{C}=\hat{x}\Gamma_{x}+\hat{y}\Gamma_{y}$, with the chiral symmetry operators along the two spatial directions $\Gamma_{x}=\sigma_z\otimes I$, $\Gamma_{y}= I \otimes \sigma_z$,
and $\tilde{\tau}$ is the total evolution time. 
For sufficiently long evolution time, $v_{2d}$ oscillates around the 2D winding number extracted from the 2D Zak phases~\cite{supp,PhysRevLett.118.076803}, providing a bulk dynamic probe to the topological invariant. 


For the experiment, we initialize the BEC at the central site of the 2D lattice, as indicated in the subfigure of Fig.~\ref{fig5}({\bf a}).
We then perform time evolutions with $\tilde{\tau}=0.6$ ms, and calculate $v_{2d}$ according to
Eq.~(\ref{2Dwinding}). Figure \ref{fig5}({\bf b}) summarizes a series of measurements with the lattice parameter $\lambda$ varying from -0.5 to 0.5.
Consistent with the theoretical prediction,
we observed that in the topological trivial phase ($\lambda <0$), the measured $v _{2d}$ oscillates around 0, corresponding to the 2D Zak phases $(0,0)$ of Fig.~\ref{fig5}({\bf a}). While for the non-trivial phase ($\lambda >0$), the measured $v _{2d}$ oscillates around 1, corresponding to the 2D Zak phases $(\pi,\pi)$ of Fig.~\ref{fig5}({\bf a}).

To match the measurement results above, we further probe the BICs in different topological phases using boundary dynamics.
We initialize the condensate at the corner site of the 2D array and measure the ensuing population evolution. In the topological trivial phase ($\lambda =-0.5$), Fig.~\ref{fig5}({\bf c}) shows that the
measured population distribution at $\tau=0.5$ ms already diffuses toward the bulk. While in the topological non-trivial phase ($\lambda =0.5$), Fig.~\ref{fig5}({\bf e}) exhibits localized population, consistent with the emergence of the BIC. At the critical point ($\lambda =0$), a fully diffusive population of the bulk state is observed, as depicted in Fig. 4(d).

{\it Discussion.}
In conclusion, we have experimentally created a 2D momentum lattice in an ultracold gas of $^{87}$Rb atoms.
Using this 2D programable platform, we have demonstrated the design and manipulation of 2D SSH model, observing the characteristic dynamics of corner and edge bound states. Facilitated by the flexible control of our setup, we engineered a time-dependent Hamiltonian to adiabatically prepare the zero-energy BICs. We further harness the bound-state dynamics and the 2D topological invariant to measure the intrinsic higher-order topological phase transition.

For future studies, it would be worthwhile to extend our present 2D lattice configuration to more complicated geometries and internal-state configurations~\cite{PRXQuantum.5.010310}. Additionally, the tuning of long-range interactions inherent to the momentum-lattice~\cite{An2021a, Xiao2021, PhysRevLett.129.103401} would also provide possibilities to study the manipulation of higher-order topological quantum matter in the strongly correlated regime~\cite{Tokura2022,Meng2023}.

\begin{acknowledgments}
{\it Acknowledgement:}
We acknowledge the support from the National Key Research and Development Program of China under Grant No.2023YFA1406703 and No. 2022YFA1404203, the National Natural Science Foundation of China under Grants Nos. 12425408, U21A20437, 12074337, and 12374479, the Fundamental Research Funds for the Central Universities under Grant No. 2021FZZX001-02  and 226-2023-00131, the China Postdoctoral Science Foundation under Grant No. 2024T170763 and GZB20240666.
\end{acknowledgments}

Z. D., H. L., and H. W., contributed equally to this work.

\bibliographystyle{apsrev4-2}
\bibliography{2Dssh}

\end{document}